\documentclass{article}
\usepackage{graphicx}
\usepackage{amsmath}

\input{tcilatex}

\begin{document}

\title{''Active'' Teleportation of a Quantum Bit}
\author{S. Giacomini, F. Sciarrino, E. Lombardi and F. De Martini \\
%EndAName
Istituto Nazionale di Fisica della Materia, Dipartimento di Fisica\\
Universit\`{a} ''La Sapienza'', Roma, 00185 Italy}
\maketitle

\begin{abstract}
We report the experimental realization of the ''active'' quantum
teleportation (QST)\ of a one-particle entangled qubit. This demonstration
completes the original QST protocol and renders it available for actual
implementation in quantum computation networks. It is accomplished by
implementing a $8m$ optical delay line and a single-photon triggered fast
Electro-Optic Pockels cell. A large value of teleportation ''fidelity'' was
attained:\ $F_{a}=(90\mp 2)\%$. Our work follows the line recently suggested
by H. W. Lee and J. Kim, Phys. Rev. A 63, 012305 (2000) and E.Knill,
R.Laflamme and G.Milburn Nature 409: 46 (2001). PACS: 03.65.Ud, 03.67.Hk,
42.50.Ar, 89.70.+c
\end{abstract}

Quantum state teleportation (QST), introduced by C. H. Bennett, G.Brassard,
C. Crepeau, R. Jozsa, A. Peres and W. Wootters came to be recognized in the
last decade as a fundamental method of quantum communication and, more
generally as one of the basic ideas of the whole field of quantum
information \cite{1}. Following the original teleportation paper and its
continuous-variables version \cite{2} an intensive experimental effort
started for the practical realization of teleportation. Quantum state
teleportation (QST) was in facts realized in a number of experiments \cite{3}%
, \cite{4} and \cite{5}. Very recently a ''qubit teleportation''\ with a
unprecedented large ''fidelity''\ $(F$ $\approx 0.95)$ has been
experimentally demonstrated by our laboratory in the context of quantum
optics by adoption of the concept of \ ''entanglement of one photon with the
vacuum'' by which each quantum superposition state, i.e. a \textit{qubit}
was physically implemented by a two dimensional subspace of Fock states of a
mode of the electromagnetic field, specifically the space spanned by the
QED\ ''vacuum'' and the 1-photon state \cite{6}. Precisely, if $A$ and $B$
represent two different modes of the field, with wavevectors (wv) $k_{A}$
and $k_{B}$ directed respectively towards two distant stations (Alice and
Bob), these ones may be linked by a non-local channel expressed by an
entangled state implying the quantum superposition of a \textit{single photon%
}, e.g. by the \textit{singlet}: $\left| \Phi \right\rangle _{singlet}$ = $%
2^{-%
%TCIMACRO{\UNICODE[m]{0xbd}}%
%BeginExpansion
{\frac12}%
%EndExpansion
}(\left| 1\right\rangle _{A}\left| 0\right\rangle _{B}-\left| 0\right\rangle
_{A}\left| 1\right\rangle _{B})$. Here the\ mode indexes 0 and 1 denote
respectively the vacuum and 1-photon Fock state population of the modes $%
k_{A},$ $k_{B}$ and the superposition state may be simply provided by an
optical \textit{beam splitter} (BS), as we shall see. The relevant
conceptual novelty introduced here consists of the fact that the \textit{%
field's modes }rather than the photons\textit{\ }associated with them are
taken as the information carriers, i.e. \textit{qubits. }Furthermore here a
proper use is made of the concept of ''single-photon nonlocality'', a
paradigm first introduced by Albert Einstein in a context close to the
formulation of the EPR\ ''paradox'' \cite{7, 8} and later elaborated by \cite
{9}, \cite{10}, \cite{11} and \cite{12}. Our scheme was designed by adapting
the methods proposed recently by Knill et.al. \cite{13} and by M.Duan et al. 
\cite{14}.

All these previous QST realizations nevertheless correspond to highly
simplified, \textit{incomplete} ''passive'' schemes by which the success of
the protocol is demonstrated indirectly by the mere detection of the
nonlocal correlations established a posteriori between the extreme stations,
Alice and Bob. These \textit{passive} realizations, indeed practically
useless and intended for mere demonstration have the advantage of avoiding
the difficult \ implementation of the final stage of QST\ i.e. of the
unitary transformations U restoring the \textit{exact }input qubit at Bob's
site under Alice's control through the QST \textit{local channel} \cite{1}.
The main problem faced here is due to the relatively long time T needed to
switch, under \textit{single-photon} excitation by the Alice's
Bell-measurement apparatus the high-voltage (HV)\ pulses driving the
electro-optic Pockels-cells (EOP)\ which implement the necessary U-unitaries
at Bob's site. Of course, in order to preserve an appreciable QST\ \textit{%
fidelity }it must be: $T\ll \tau $, being $\tau $ the characteristic time of
the de-coherence process affecting the \textit{nonlocal channel}, i.e. the
one that dephases the corresponding entangled $\left| \Phi \right\rangle
_{singlet}$ \cite{1}. The present work realizes for the first time the 
\textit{complete,} i.e.\textit{\ active }qubit\ teleportation process by
completing the corresponding optical scheme according to the standard QST
protocol \cite{1}. As a basic\ qubit-QST scheme, the quoted vacuum-1 photon
configuration was adopted \cite{6}.

The experimental set-up, Figure 1, can be somewhat considered to be the
''folded'' configuration of the one reported in \cite{6}. The significant
changes consisted of the addition of the optical delay line $(DL)$\ and of a
different measurement apparatus at Bob's site. Let us briefly outline the
details of the apparatus. A nonlinear (NL) $LiIO_{3}$ crystal slab, 1.5 mm
thick with parallel anti-reflection coated faces, cut for Type I
phase-matching was excited by a single mode UV cw argon laser with
wavelength (wl) $\lambda _{p}=$ $363.8nm$ and with an average power $\simeq
100mW$. The two photons belonging to each correlated pair emitted by
spontaneous parametric down conversion (SPDC) with the same linear vertical
(V) polarization had equal wavelengths (wl) $\lambda =727,6nm$ and were
spatially selected by equal pinholes with diameter $0.5mm$ placed at a
distance of $50cm$ from the NL\ crystal. Of course, the product state
condition of each SPDC pair, $\left| \Phi \right\rangle _{out}=$ $\left|
1\right\rangle _{A}\otimes \left| 1\right\rangle _{S}$ did not imply any
mutual nonlocal correlation between the particles. Indeed for our purpose
the particles could have been emitted by two totally independent sources. By
two beam splitters devices $BS$ and $BS_{S}$, each composed by a combination
of a $\lambda /2$ plate + a calcite crystal (C) and inserted on the output
modes of the SPDC\ source, $\left| \Phi \right\rangle _{out}$ was
transformed into the product of two entangled states, $\left| \Phi
\right\rangle _{singlet}=2^{-%
%TCIMACRO{\UNICODE[m]{0xbd}}%
%BeginExpansion
{\frac12}%
%EndExpansion
}(\left| 1\right\rangle _{A}\left| 0\right\rangle _{B}-\left| 0\right\rangle
_{A}\left| 1\right\rangle _{B})$ and $\left| \Psi \right\rangle _{S%
\widetilde{a}}$= $(\overline{\alpha }\left| 1\right\rangle _{S}\left|
0\right\rangle _{\widetilde{a}}+\overline{\beta }\left| 0\right\rangle
_{S}\left| 1\right\rangle _{\widetilde{a}})$ with $\overline{\alpha }^{2}+%
\overline{\beta }^{2}=1$, defined over the two pairs of the $BS^{\prime }s$
output modes $(k_{A},k_{B})$ and $(k_{S},k_{\widetilde{a}})$, respectively.
Precisely, one photon of the SPDC\ pair excited the output modes $k_{A}$, $%
k_{B}$ of the symmetrical, i.e. 50:50 beam splitter $BS$ by the singlet
state $\left| \Phi \right\rangle _{singlet}$ which provided the nonlocal
teleportation channel. The output modes $k_{S},k_{\widetilde{a}}$ of the
other, \textit{variable}\ beam splitter $BS_{S}$ were excited by the
entangled state $\left| \Psi \right\rangle _{S\widetilde{a}}$ giving rise to
the \textit{local} realization on the output mode $k_{S}$ of the \textit{%
unknown} qubit to be teleported: $\left| \Psi \right\rangle _{in}=(\alpha
\left| 0\right\rangle _{S}+\beta \left| 1\right\rangle _{S})$. This
corresponded to the local realization on the mode $k_{\widetilde{a}}$ of a
related ''ancilla'' state $\left| \Psi \right\rangle _{\widetilde{a}%
}=(\gamma \left| 0\right\rangle _{\widetilde{a}}+\delta \left|
1\right\rangle _{\widetilde{a}})$ that will be adopted here for the
verification of the QST\ success, as we shall see. Most important, the
''ancilla''\ provided in our system the synchronizing ''clock'' signal that
has been recognized to be a necessary\ ingredient of every quantum computing
network when single-photon \textit{qubits} and \textit{e-bits} are involved 
\cite{6,13,14}. At Alice's site, the modes $k_{A}$ and $k_{S}$ were linearly
superimposed by a common 50:50 beam splitter $BS_{A}$ whose output modes $%
k_{1}$, $k_{2}$ excited the Bell-measurement apparatus, consisting of the
detectors $D_{1},D_{2}.$ Micrometric changes of the mutual phase $\varphi $
of the interfering $k_{S}$ and $k_{A}$ modes were obtained by a
piezoelectrically driven mirror placed at the exit of $BS_{S}$. All
detectors in the experiment were Si-avalanche EG\&G-SPCM200 modules having
equal quantum efficiencies $QE\approx $ $0.45$. The beams were filtered
before detection by interference-filters within a bandwidth $\Delta \lambda
=6nm$.

The apparatus at the Alice site, consisting of $BS_{A},D_{1},D_{2}$
performed a standard Bell-state measurement with 50\% efficiency. Precisely,
the Bell states $\left| \Psi ^{3}\right\rangle _{SA}=2^{-%
%TCIMACRO{\UNICODE[m]{0xbd}}%
%BeginExpansion
{\frac12}%
%EndExpansion
}(\left| 0\right\rangle _{S}\left| 1\right\rangle _{A}-\left| 1\right\rangle
_{S}\left| 0\right\rangle _{A})$ and $\left| \Psi ^{4}\right\rangle
_{SA}=2^{-%
%TCIMACRO{\UNICODE[m]{0xbd}}%
%BeginExpansion
{\frac12}%
%EndExpansion
}(\left| 0\right\rangle _{S}\left| 1\right\rangle _{A}+\left| 1\right\rangle
_{S}\left| 0\right\rangle _{A})$ implying the teleportation of a
single-photon qubit, could be singled out by our scheme \cite{6}. The other
two Bell states involving measurements at Alice's of either zero or two
photons $\left| \Psi ^{1}\right\rangle _{SA}${}$=${}$\left| 0\right\rangle
_{S}\left| 0\right\rangle _{A}$, $\left| \Psi ^{2}\right\rangle _{SA}$=$%
\left| 1\right\rangle _{S}\left| 1\right\rangle _{A}$ were not identified by
the apparatus, as expected for any linear detection system \cite{15}.
Furthermore, it could be easily checked by carrying out formally the product 
$\left| \Psi \right\rangle _{total}=\left| \Psi \right\rangle _{in}\left|
\Phi \right\rangle _{singlet}$ that a single photo-detection, a ''click'' by 
$D_{1}$, i.e. the realization over the $BS_{B}$ output mode set $%
(k_{1},k_{2})$ of the state $\left| 1\right\rangle _{1}\left| 0\right\rangle
_{2}=\left| \Psi ^{3}\right\rangle _{SA}$, implied the realization at Bob's
site of the state $(\alpha \left| 0\right\rangle _{B}+\beta \left|
1\right\rangle _{B})$, i.e. of the \textit{exact} teleported copy of the
input state $\left| \Psi \right\rangle _{in}$. Alternatively, a click at $%
D_{2}$ expressing the realization of the state $\left| 0\right\rangle
_{1}\left| 1\right\rangle _{2}=\left| \Psi ^{4}\right\rangle _{SA}$ implied
the realization at Bob's site of the state $(\alpha \left| 0\right\rangle
_{B}-\beta \left| 1\right\rangle _{B})=\sigma _{z}\left| \Psi \right\rangle
_{in}$. Since $\sigma _{z}^{2}=I$, the complete QST protocol could be
achieved in our experiment by allowing the direct activation by $D_{2}$ of
an electro-optic (EO) device performing at Bob's site the unitary
transformation $U\equiv \sigma _{z}$. This transformation was implemented in
the experiment by a $LiNbO_{3}$ high-voltage micro Pockels Cell (EOP)\ made
by Shangai Institute of Ceramics with \TEXTsymbol{<}1 nsec risetime. The EOP 
$\lambda /2$ voltage, i.e. leading to a $\lambda /2$ EO-induced\ phase shift
of the single-photon state $\beta \left| 1\right\rangle $ respect to the
(phase-insensitive) vacuum state $\alpha \left| 0\right\rangle _{B}$
appearing in the expression of $(\sigma _{z}\left| \Psi \right\rangle _{in})$%
, was $V_{\lambda /2}=1.4kV$. The difficult problem of realizing an fast
electronic circuit transforming each small ($\simeq $1 mV) Si-avalanche
photodetection signal into a calibrated fast pulse in the kV range was
solved by a single linear amplifier chip (LM9696) exciting directly a single
chain of six fast avalanche transistors (Zetex ZXT413:\ cfr: Fig.1, inset).
The overall risetime achieved by the device was $=22$ nsec. This
corresponded to the minimum delay we must impart to the teleported single
photon\ state $(\sigma _{z}\left| \Psi \right\rangle _{in})$ before entering
EOP\ in order to be transformed into the wanted state $\left| \Psi
\right\rangle _{in}$. The optical $DL$, $8m$ long and corresponding to a
delay $\Delta T=24n\sec $ \TEXTsymbol{>}$T_{r}$, was realized by multiple
reflections by three dielectric coated plane mirrors and by two
anti-reflection coated high-quality lenses: $L_{1}$ with focal length 1.5 m
and $L_{2}$ with f.l.=0.3 m (cfr. Fig. 1).

An efficient test of the success of the \textit{active} qubit-QST operation
was carried out by a ''passive'' interference procedure involving the
synchronizing \textit{clock} state, i.e. the ''ancilla'' state associated to
the mode $k_{\widetilde{a}}$. The beam carrying the ''ancilla'' state $%
\left| \Psi \right\rangle _{\widetilde{a}}$was injected into the main
optical delay line by means of a polarizing beam splitter (PBS) and was made
to fully spatially overlap the main teleportation beam, i.e. carrying the
states $\left| \Psi \right\rangle _{in}$ and $(\sigma _{z}\left| \Psi
\right\rangle _{in})$, but with \textit{orthogonal} polarization respect to
these ones. Precisely, the EO\ crystal of the Pockels cell was oriented in a
direction such that the efficiency of the EO phase-shifting operation was
maximum for the (V)\ polarization of the main teleported beams and zero for
the horizontal (H)\ polarization of the ''ancilla'' beam.

At the exit of the EOP, the orthogonal linear polarizations of the two
(teleported+ancilla) beams were first rotated by the same amount by means of
a $\lambda /4$ plate, then linearly mixed by a polarizing beam splitter $%
PBS_{B\text{ }}$ and finally measured by the couple of verification
detectors: $D\QATOP{\ast }{1},D\QATOP{\ast }{2}$. The combination $(\lambda
/4$ plate+$PBS_{B})$ provided the \textit{variable} beam-splitting action $%
(BS_{B})$ at Bob's site needed to implement the ''passive'' verification of
the QST success through coincidence measurements involving the two detector
sets $(D_{1},D_{2})$ and $(D\QATOP{\ast }{1},D\QATOP{\ast }{2})$. Assume for
simplicity and with no loss of generality that both $BS_{S\text{ }}$and $%
BS_{B}$ were set in the \textit{symmetrical} condition: i.e. with equal
reflectivity and transmittivity parameters: $\left| r\right| ^{2}=\left|
t\right| ^{2}=%
%TCIMACRO{\UNICODE[m]{0xbd}}%
%BeginExpansion
{\frac12}%
%EndExpansion
$. Let us vary\ the position $X=$ $(2)^{-3/2}\lambda \varphi /\pi $ of the
mirror at the exit of $BS_{B}.$ By straightforward calculation the phase
shift $\varphi $ induced on the measured fields is found to affect the
coincidence counts rates according to the following expressions \cite{6}: 
\begin{equation*}
(D_{1}-D_{2}^{\ast })=(D_{2}-D_{1}^{\ast })=%
%TCIMACRO{\UNICODE[m]{0xbd}}%
%BeginExpansion
{\frac12}%
%EndExpansion
\cos ^{2}\frac{\varphi }{2};(D_{1}-D_{1}^{\ast })=(D_{2}-D_{2}^{\ast })=%
%TCIMACRO{\UNICODE[m]{0xbd}}%
%BeginExpansion
{\frac12}%
%EndExpansion
\sin ^{2}\frac{\varphi }{2},
\end{equation*}
where $(D_{i}-D_{j}^{\ast })$ , $i,j=1,2$, expresses the probability of a
coincidence detected by the pair $D_{i}$, $D_{j}^{\ast }$ in correspondence
with the realization at Alice's site either of the state $\left| \Psi
^{3}\right\rangle _{SA}$, for $i\neq j$, or of the state: $\left| \Psi
^{4}\right\rangle _{SA}$ for $i=j$. The realization of these states implied
the corresponding realization at Bob's site of the teleported state $\left|
\Psi \right\rangle _{in}$ or of the state $(\sigma _{z}\left| \Psi
\right\rangle _{in})$, as said. \ The upper and lower plots shown in Figure
2 correspond to the actual realization of these states. Precisely, the upper
plot shows the experimental coincidence data corresponding to the direct
realization at Bob's site of the teleported $\left| \Psi \right\rangle _{in}$%
. The data of the lower plot expressed by full circles correspond to the
realization at Bob's site of the state $(\sigma _{z}\left| \Psi
\right\rangle _{in})$ upon \textit{inhibition} of the EOP action. However
the data expressed by open circles, taken by previous EOP\ activation,
correspond to the realization of the EOP-transformed state $U(\sigma
_{z}\left| \Psi \right\rangle _{in})=\left| \Psi \right\rangle _{in}$, as
shown by comparison with the upper plot. In summary, the results show that
allowing the automatic actuation of the EOP-switch by the Alice's detector $%
D_{2}$ results in the actual teleportation of the unknown input qubit $%
\left| \Psi \right\rangle _{in}$ \textit{whenever }one of the detectors $%
D_{1}$, $D_{2}$ clicks, i.e. whenever a single-photon is detected at Alice's
site and another single-photon is detected at Bob's site. Note that the
above QST verification procedure involving the ancilla mode $k_{\widetilde{a}%
}$ enabled a nearly \textit{noise-free} teleportation procedure. Indeed, if
no photons were detected at Alice'site, i.e. by $D_{1}$and/or $D_{2}$, while
photons were detected at Bob's site by $D_{1}^{\ast }$ and/or $D_{2}^{\ast }$
we could conclude that the ''idle'' Bell state $\left| \Psi
^{1}\right\rangle _{SA}$ was created. If on the contrary no photons were
detected at Bob'site while photons were detected at Alice's site, we could
conclude that the other ''idle'' Bell state $\left| \Psi ^{2}\right\rangle
_{SA}$ was realized. The data collected in correspondence with these
''idle'' events were automatically\ discarded by the electronic coincidence
circuit.

Note, interestingly that the presence of the delay-line did not impair
substantially the value of the QST ''fidelity'', as determined by \ the
''visibility''\ $V$\ of the interference plots of Figure 2:\ $F\equiv
\left\langle \Phi _{in}\right| \rho _{out}\left| \Phi _{in}\right\rangle
=(1+V)/2$. It was found that the very high value $F=(95.3\pm 0.6)\%$
previously attained by the \textit{passive} method was reduced by the
present $DL$ de-coherence to the figure: $F_{a}=(90\pm 2)\%$ , still largely
overcoming the limit value implied by genuine quantum teleportation \cite{3}%
. All these results fully implement within the framework of the vacuum-1
photon QST the original \textit{complete}, i.e. \textit{active}
teleportation protocol \cite{1,6}. This method can be easily extended to the
other common qubit QST configurations \cite{3,4}, where in general two
different unitary transformations, and then two different EOP\ devices are
required: $U_{j}(\sigma _{x},\sigma _{y}),$ $j=1,2$.

Because of the prospective key relevance of the quantum teleportation
protocol at the core of many important logic networks \cite{16}, the present
demonstration is expected to represent a substantial step forward towards
the actual implementation of complex multiple qubit gates in the domain of
linear optics quantum computation \cite{13,14}. The recent literature shows
that the field is indeed moving fast in that direction \cite{17}.

We acknowledge useful and lively conversations with Sandu Popescu. We are
also greatly indebted with the FET European Network on Quantum Information
and Communication (Contract IST-2000-29681:ATESIT) and with M.U.R.S.T. for
funding.

\centerline{\bf Figure Captions}

\vskip 8mm

\parindent=0pt

\parskip=3mm

1. Experimental apparatus realizing the ''active'' quantum state
teleportation protocol (QST). The cross close to the center of the optical
delay line indicates the approximate position of the focal region of the
lens $L_{1}$. INSET: Diagram of the fast electronic switch of the
Electro-Optic Pockels cell (EOP) implementing the unitary transformation:\ $%
U\equiv \sigma _{z}$.

2. Interferometric fringe patterns due to coincidence experiments involving
different pairs of detectors within the \textit{active} QST verification
procedure upon variation of the phase $\varphi $ of the measured fields.
Upper plot: pattern related to the \textit{exact} teleportation of the input
state:\ $\left| \Psi \right\rangle _{in}$. Lower plot, full circles:
teleportation of the state $(\sigma _{z}\left| \Psi \right\rangle _{in})$
with inhibition of the EOP operation. Lower plot, open circles: result of
the transformation $U(\sigma _{z}\left| \Psi \right\rangle _{in})=\left|
\Psi \right\rangle _{in}$ induced by the EOP activation by the Alice's
detector $D_{2}$.\newline

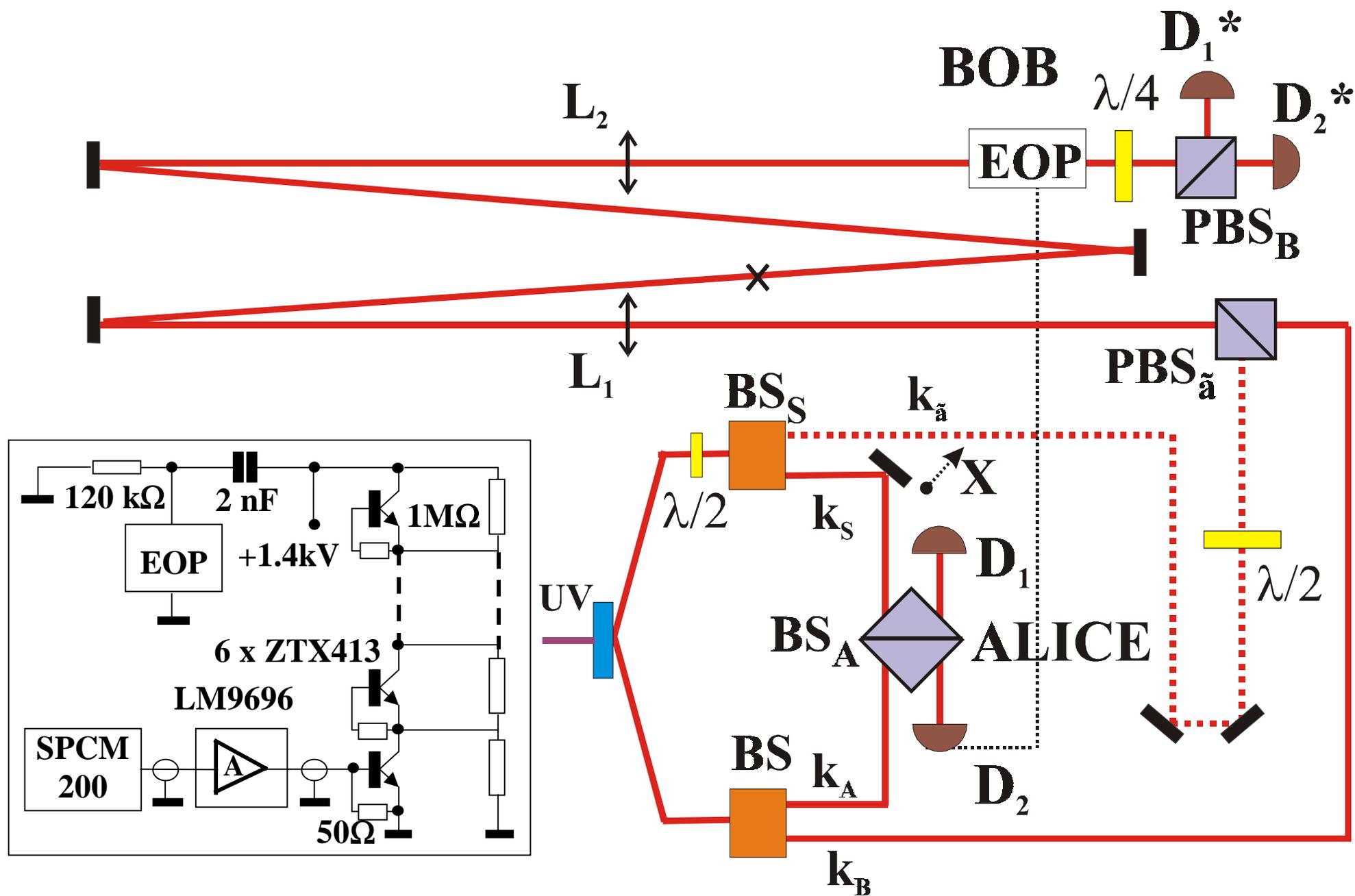
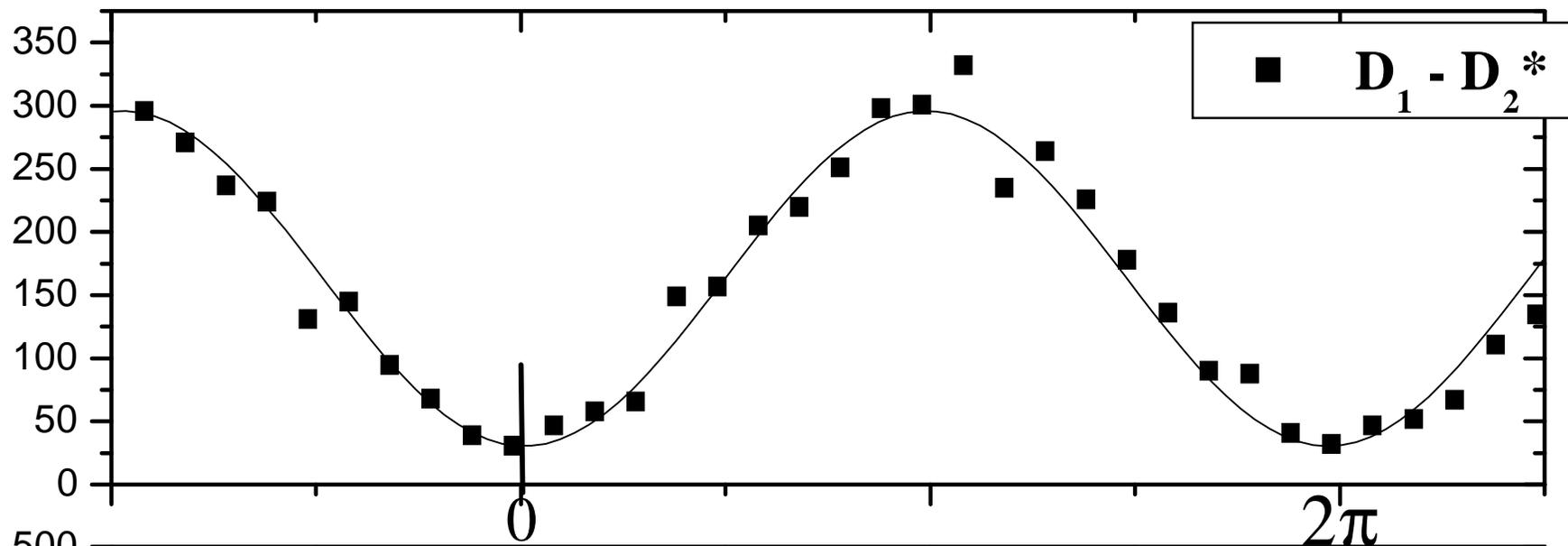
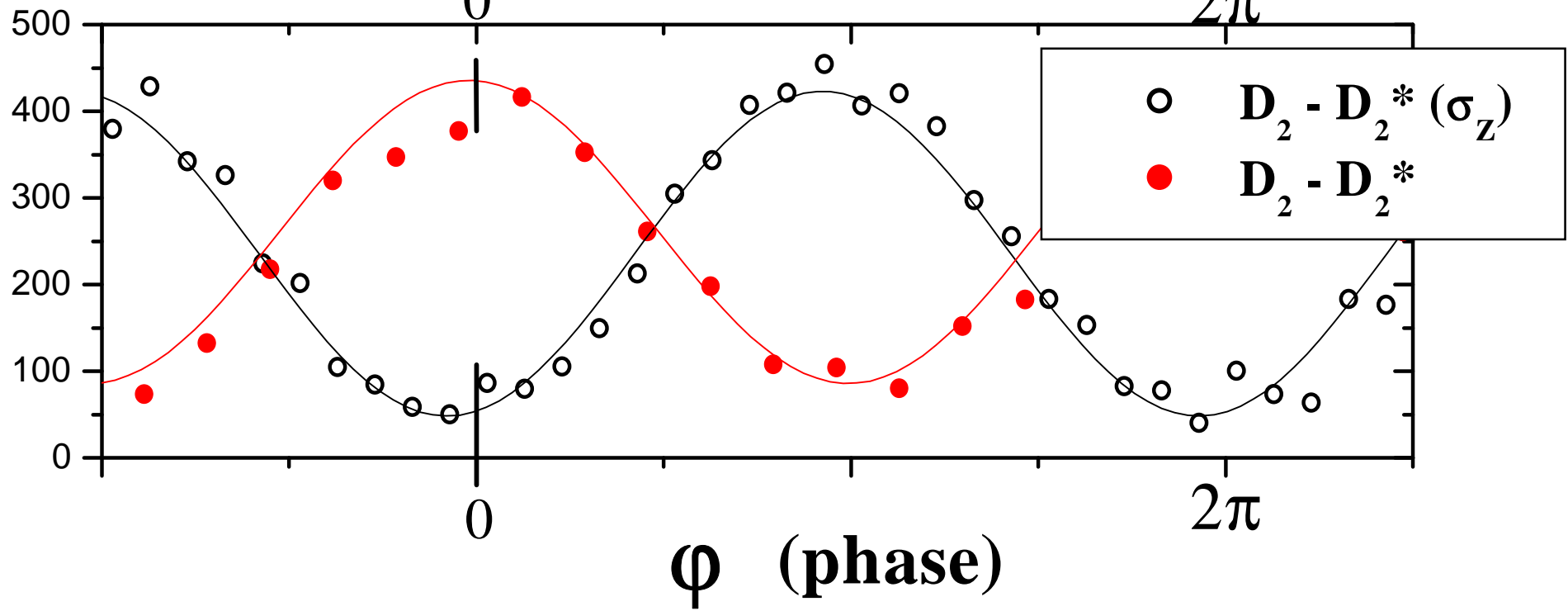
\end{document}